# Cascaded Reconstruction Network for Compressive image sensing

**Yahan Wang, Huihui Bai, Lijun Zhao, Yao Zhao**
Institute of Information Science, Beijing Jiaotong University
Beijing, 100044 - China

***Abstract***

The theory of compressed sensing (CS) has been successfully applied to image compression in the past few years, whose traditional iterative reconstruction algorithm is time-consuming. Fortunately, it has been reported deep learning-based CS reconstruction algorithms could greatly reduce the computational complexity. In this paper, we propose two efficient structures of cascaded reconstruction networks corresponding to two different sampling methods in CS process. The first reconstruction network is a compatibly sampling reconstruction network (CSRNet), which recovers an image from its compressively sensed measurement sampled by a traditional random matrix. In CSRNet, deep reconstruction network module obtains an initial image with acceptable quality, which can be further improved by residual network module based on convolutional neural network. The second reconstruction network is adaptively sampling reconstruction network (ASRNet), by matching automatically sampling module with corresponding residual reconstruction module. The experimental results have shown that the proposed two reconstruction networks outperform several state-of-the-art compressive sensing reconstruction algorithms. Meanwhile, the proposed ASRNet can achieve more than 1 dB gain, as compared with the CSRNet.

***Keywords:*** Compressive sensing, Sampling net, Reconstruction net, CSRNet, ASRNet

# 1. Introduction

In the traditional Nyquist sampling theory, the sampling rate must be at least twice of the signal bandwidth in order to reconstruct the original signal losslessly. On the contrary, compressive sensing (CS) theory is a signal acquisition paradigm, which can sample a signal at sub-Nyquist rates but realize the high-quality recovery [1]. Later, Gan et al. proposed block compresses sensing to reduce the algorithm's computational complexity to avoid directly applying CS on images with large size [2]. Due to CS's excellent performance on sampling, CS has already been widely used in a great deal of fields, such as communication, signal processing, etc.

In the past decades, CS theory has advanced considerably, especially in the development of reconstruction algorithms [3-10]. Compressive sensing recovery aims to recover the original signal $x \in R^{n \times 1}$ from the compressive sensing measurement $y \in R^{m \times 1} (m << n)$. The CS measurement is obtained by $y = \Phi x$, where $\Phi \in R^{m \times n}$ is a CS measurement matrix. The process of reconstruction is highly ill-posed, because there exist more than one solutions $x \in R^{n \times 1}$ that can generate the same CS measurement $y$. To solve this problem, the early recovery algorithms always assume the original image signal has the property of $l_p$-norm $(0 \leq p \leq 1)$ sparsity. Based on this assumption, several iterative reconstruction algorithms have been explored, such as orthogonal matching pursuit (OMP) [3] and approximate message passing(AMP) [4]. Distinctively, the extension of the AMP, Denoising-based AMP (D-AMP) [5] employs denoising algorithms for CS recovery and can get a high performance for nature images. Furthermore, many works incorporate prior knowledge of the original image signals, such as total variation sparsity prior [6] and KSVD [7], into CS recovery framework, which can improve the CS reconstruction performance. Particularly, TVAL3[8] combines augmented Lagrangian method with total variation regularization, which is also perfect CS image reconstruction algorithm. However, almost all these recovery algorithms require to solve an optimization problem. Most of those algorithms need hundreds of iterations, which inevitably leads to high computational cost and becomes the obstacle for the applications of CS.

In recent years, some deep learning based methods have been introduced into the low-level problems and get excellent performance, such as image super-resolution [11-12], image artifact removal [13], and CS image recovery [14-17]. Recently, some deep network-based algorithms for CS image recovery have been proposed. ReconNet, is proposed in [14], which takes CS measurement of image patch as input and outputs its corresponding image reconstruction. Especially, for patch-based CS measurement, ReconNet, inspired of SRCNN [11], can retain rich semantic contents at low measurement rate as compared to the traditional methods. In [15], a framework is proposed to recover images from CS measurements without the need to divide images into small blocks, but there is no competitive advantage for the performance of the reconstruction compared with other algorithms. In [16-17], the residual convolutional neural network is introduced in the image reconstruction for compressive sensing, which can preserve some information in previous layers and also can improve the convergence rate and accelerate the training process. Different from the optimization-based CS recovery methods, the neural network-based methods often directly learn the inverse mapping from the CS measurement domain to original image domain. As a result, it effectively avoids expensive computation, and achieves a promising image reconstruction performance.

In this paper, two different cascaded reconstruction networks are proposed to meet different sampling methods. Firstly, we propose a compatibly sampling reconstruction network (CSRNet), which is employed to reconstruct high-quality images from compressively sensed measurements sampled by a random sampling matrix. In CSRNet, deep reconstruction network module can obtain an initial image with acceptable quality, which can be further improved by residual network module based on convolutional neural network. Secondly, in order to improve the sampling efficiency of CS, an automatically sampling module is designed, which has a fully-connected layer to learn a sampling matrix automatically. In addition, the residual reconstruction module is presented, which can match the sampling module. Both the sampling module and its matching residual reconstruction module form a complete compressive sensing image reconstruction network, named ASRNet. As compared with CSRNet, ASRNet can achieve more than 1 dB gain. The experimental results demonstrate the proposed networks outperform several state-of-the-art iterative reconstruction algorithms and deep-learning based approaches in objective and subjective quality.

The rest of this paper is organized as follows. In Section 2, two novel networks are proposed for different sampling methods. In Section 3, the performance of the proposed networks is examined. We conclude the paper in Section 4.
.

## 2. Proposed networks

In this section, we describe the proposed two networks CSRNet in **Fig. 1** and ASRNet in **Fig. 4**. The first network, CSRNet, is designed to reconstruct image from the CS measurement sampled by a random matrix. The second one is a complete compressive sensing image reconstruction network, ASRNet, consisting of both sampling and recovery module. Here, our sampling module contains only one fully-connected layer(FC), which is more powerful to imitate traditional Block-CS sampling process.

### 2.1 CSRNet

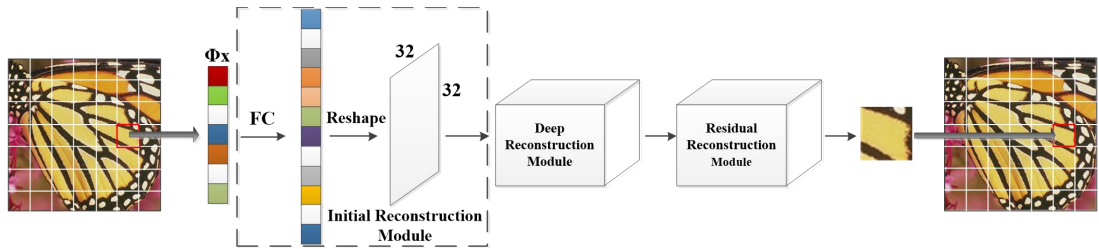

**Fig.1.** The framework of CSRNet

Our proposed CSRNet consists of three modules, initial reconstruction module, deep reconstruction module and residual reconstruction module. The initial reconstruction module takes the CS measurement $y$ as input and outputs a $B×B$ sized preliminary reconstructed image. As shown in the Fig.1, the deep reconstruction module takes the preliminary reconstructed image as input and outputs a same sized image. The deep reconstruction module contains three convolutional layers, shown in **Fig.2**. The first layer generates 64 feature map with $11×11$ kernel. The second layer uses kernel of size *1×1* and generates 32 feature maps. And the third layer produces one feature map with *7×7* kernel, which is the output of this module. All the convolutional layers have the same stride of 1, without pooling operation, and appropriate zero padding is used to keep the feature map size constant in all layers. Each convolutional layer is followed by a ReLU layer except the last convolutional layer. Here, deep reconstruction network module can obtain an initial image with acceptable quality, which is more suitable to residual network module than cascaded residual network module [16]. The residual reconstruction network has the similar architecture as the deep reconstruction network, shown in **Fig.3**, which learns the residual information between the input data and the ground truth. In our model, we set *B=32*.

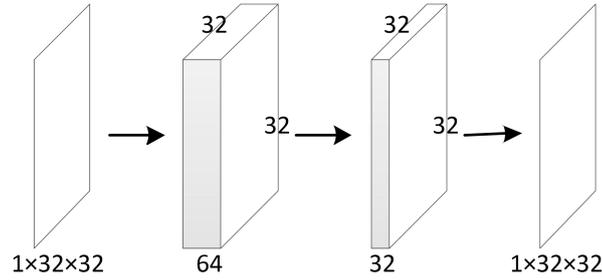

**Fig.2.** The framework of deep reconstruction module

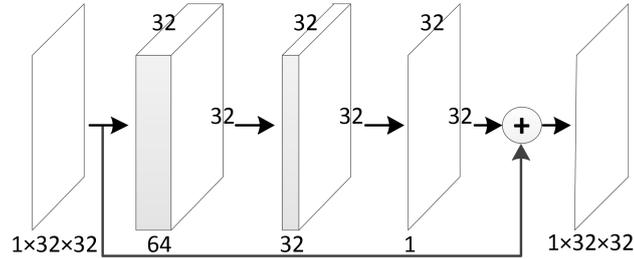

**Fig.3.** The framework of residual reconstruction module

In order to train our CSRNet, we need CS measurements corresponding to each of the extracted patches. For a given measurement rate, we construct a measurement matrix, $\Phi_B$ by first generating a random Gaussian matrix of appropriate size, followed by orthonormalizing its rows. Then, we apply $y_i = \Phi_B x_{vec-i}$ to obtain the set of CS measurements, where $x_{vec-i}$ is the vectorized version of an image patch $x_i$. Thus, an input-label pair in the training set can be represented as $\{y_i, x_i\}_i^N$. The loss function is the average reconstruction error over all the training image blocks, given by

$$L(\{W_1, W_2, W_3\}) = \frac{1}{N} \sum_{i=1}^{N} \left\| f_3(f_2(f_1(y_i, \{W_1\}), \{W_2\}), \{W_3\}) - x_i \right\|^2 \tag{1}$$

where *N* is the total number of image patches in the training dataset, $x_i$ is the $i^{th}$ patch and $y_i$ is the corresponding CS measurement. The initial reconstruction mapping ,the deep reconstruction mapping and the residual reconstruction mapping are represented as $f_1$, $f_2$, and, $f_3$ respectively. In addition, $\{W_1, W_2, W_3\}$ are the network parameters which can be obtained in the training.

**2.2 ASRNet**

Our proposed ASRNet contains three modules, sampling module, initial reconstruction module and residual reconstruction module. In the sampling module, we use a fully-connected layer to imitate the traditional compressed sampling process. And the process of compressed sampling is expressed as $y_i = \Phi_B x_i$ in traditional Block-CS. If the image is divided into $B \times B$ blocks, the input of the fully-connected layer is a $B^2 \times 1$ vector. For the sampling ratio $\alpha$, we can obtain $n_B = B^2 \times \alpha$ sampling measurements.

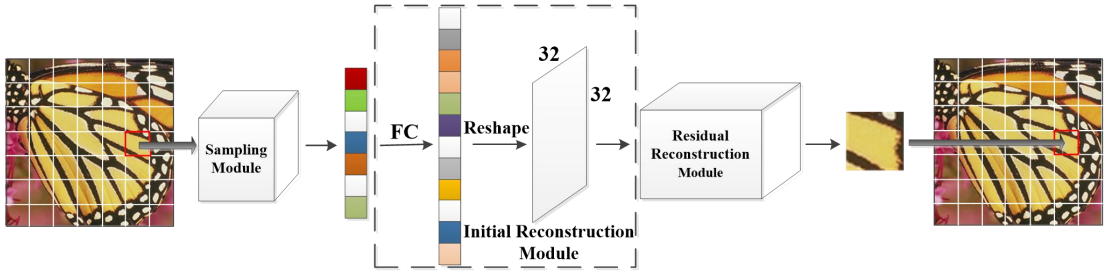

**Fig.4.** The framework of ASRNet

The initial reconstruction module and residual reconstruction module are matching with the sampling module. The initial reconstruction module takes those sampling measurements as input and outputs a $B \times B$ sized preliminary reconstructed image. Similar to sampling module, we also use a fully-connected layer to imitate the traditional initial reconstruction process, which can be presented by $\tilde{x}_j = \tilde{\Phi}_B y_j$. In our design, the $\tilde{\Phi}_B$ can be learned automatically instead of computing by the complicated MMSE linear estimation. The residual reconstruction module is similar as the residual reconstruction module in CSRNet, shown in **Fig.3**. The output of the residual reconstruction module is the final output of the network.

Given the original image $x_i$, our goal is to obtain the highly compressed measurement $y_i$ with the compressed sampling module, and then accurately recover it to the original image $x_i$ with reconstruction module. Since the sampling module, the initial reconstruction module and the residual reconstruction module form an end-to-end network, they can be trained together and do not need to be concerned with what the compressed measurement is in training. Therefore, the input and the label are all the image itself for training our ASRNet. Following most of deep learning based image restoration methods, the mean square error is adopted as the cost function of our network. The optimization objective is represented as

$$L(\{W_4, W_5, W_6\}) = \frac{1}{N} \sum_{i=1}^{N} \| f_6(f_5(f_4(x_i, \{W_4\}), \{W_5\}), \{W_6\}) - x_i \|^2 \quad (2)$$

Where $\{W_4, W_5, W_6\}$ are the network parameters needed to be trained, $f_4$ is the sampling, $f_5$ and $f_6$ correspond the initial mapping and residual reconstruction mapping respectively. It should be noted that we train the compressed sampling network and the reconstruction network together, but they can be used independently.

## 3. Experimental results

In this section, we evaluate the performance of the proposed methods for CS reconstruction. We will firstly introduce the details during our training and testing. Then, we show the quantitative and qualitative comparisons with four state-of-the-art methods.

### 3.1 Training

The dataset used in our training is the set of 91 images in the [14]. The set5 from [14] constitutes to be our validation set. We only use the luminance component of the images. We uniformly extract patches of size 32×32 from these images with a stride equal 14 for training and 21 for validation to form the training dataset of 22144 patches and the validation dataset which contains 1112 patches. Both CSRNet and ASRNet use the same dataset. We train the proposed networks with different measurement rates (MR)=0.25,0.10, 0.04 and 0.01. The Caffe is used to train the proposed model.

### 3.2 Comparison with existing methods

#### 3.2.1 Objective quality comparisons

Our proposed algorithm is compared with four representative CS recovery methods, TVAL3[8], D-AMP [5], ReconNet [14] and DR²-Net [16]. The first two belong to traditional optimization-based methods, while the last two are recent network-based methods. For the simulated data in our experiments, we evaluate the proposed methods on the same test images as in [14], which consists of 11 grayscale images. Here, nine images have size of 256×256 and two images are 512×512. We compute the PSNR value for total 11 images, and the results are shown in **Table 1**. We use the BM3D [18] as the denoiser to remove the artifacts resulting due to patch processing. From **Table 1**, it can be found that our proposed CSRNet and ASRNet outperform other algorithms under different measurement rates. It is obvious to see that ASRNet can achieve more than 1 dB gain, as compared with CSRNet.

**Table 1**. PSNR valves in dB for testing image by different algorithms at the ratio MR=0.25,0.1,0.04 and 0.01. "MEAN PSNR" is the mean PSNR values among all 11 testing images.

| Image Name | Algorithm | PSNR(without applying BM3D/with using BM3D) | | | |
|---|---|---|---|---|---|
| | | MR=0.25 | MR=0.10 | MR=0.04 | MR=0.01 |
| Barbara | TVAL3 | 24.19 / 24.20 | 21.88 / 22.21 | 18.98 / 18.98 | 11.94 / 11.96 |
| | D-AMP | 25.08 / 25.96 | 21.23 / 21.23 | 16.37 / 16.37 | 5.48 / 5.48 |
| | ReconNet | 23.25 / 23.52 | 21.89 / 22.50 | 20.38 / 21.02 | 18.61 / 19.08 |
| | DR²-Net | 25.77 / 25.99 | 22.69 / 22.82 | 20.70 / 21.30 | 18.65 / 19.10 |
| | CSRNet | 26.17 / 26.30 | 22.94 / 22.95 | 21.24 / 21.47 | 19.10 / 19.17 |
| | ASRNet | 26.30 / 26.43 | 24.34 / 24.35 | 23.48 / 23.54 | 21.40 / 21.52 |
| Fingerprint | TVAL3 | 22.70 / 22.71 | 18.69 / 18.70 | 16.04 / 16.05 | 10.35 / 10.37 |
| | D-AMP | 25.17 / 23.87 | 17.15 / 16.88 | 13.82 / 14.00 | 4.66 / 4.73 |
| | ReconNet | 25.57 / 25.13 | 20.75 / 20.97 | 16.91 / 16.96 | 14.82 / 14.88 |
| | DR²-Net | 27.65 / 27.75 | 22.03 / 22.45 | 17.40 / 17.47 | 14.73 / 14.95 |
| | CSRNet | 27.20 / 27.44 | 21.64 / 21.90 | 17.49 / 17.55 | 15.09 / 15.18 |
| | ASRNet | 28.82 / 29.23 | 26.25 / 26.83 | 20.98 / 21.45 | 16.20 / 16.21 |
| Flinstones | TVAL3 | 24.05 / 24.07 | 18.88 / 18.92 | 14.88 / 14.91 | 9.75 / 9.77 |
| | D-AMP | 25.02 / 24.45 | 16.94 / 16.82 | 12.93 / 13.09 | 4.33 / 4.34 |
| | ReconNet | 22.45 / 22.59 | 18.92 / 19.18 | 16.30 / 16.56 | 13.96 / 14.08 |
| | DR²-Net | 26.19 / 26.77 | 21.09 / 21.46 | 16.93 / 17.05 | 14.01 / 14.18 |
| | CSRNet | 25.36 / 25.37 | 20.52 / 20.82 | 17.23 / 17.35 | 14.26 / 14.46 |
| | ASRNet | 26.93 / 27.40 | 24.01 / 24.56 | 19.78 / 20.08 | 16.30 / 16.39 |
| Lena | TVAL3 | 28.67 / 28.71 | 24.16 / 24.18 | 19.46 / 19.47 | 11.87 / 11.89 |
| | D-AMP | 28.00 / 27.41 | 22.51 / 22.47 | 16.52 / 16.86 | 5.73 / 5.96 |
| | ReconNet | 26.54 / 26.53 | 23.83 / 24.47 | 21.28 / 21.82 | 17.87 / 18.05 |
| | DR²-Net | 29.42 / 29.63 | 25.39 / 25.77 | 22.13 / 22.73 | 17.97 / 18.40 |
| | CSRNet | 29.50 / 29.70 | 25.72 / 25.97 | 22.89 / 23.14 | 19.09 / 19.14 |
| | ASRNet | 30.65 / 30.89 | 28.54 / 28.78 | 25.74 / 25.93 | 21.74 / 21.93 |
| Monarch | TVAL3 | 27.77 / 27.77 | 21.16 / 21.25 | 16.73 / 16.73 | 11.09 / 11.11 |
| | D-AMP | 26.39 / 26.55 | 19.00 / 18.96 | 14.57 / 14.57 | 6.20 / 6.20 |
| | ReconNet | 24.31 / 25.06 | 21.10 / 21.51 | 18.19 / 18.32 | 15.39 / 15.49 |
| | DR²-Net | 27.95 / 28.31 | 23.10 / 23.56 | 18.93 / 19.23 | 15.33 / 15.50 |
| | CSRNet | 27.87 / 28.19 | 22.99 / 23.25 | 19.39 / 19.57 | 15.42 / 15.46 |
| | ASRNet | 29.29 / 29.60 | 27.17 / 27.50 | 23.23 / 23.49 | 17.74 / 17.85 |
| Parrot | TVAL3 | 27.17 / 27.24 | 23.13 / 23.16 | 18.88 / 18.90 | 11.44 / 11.46 |
| | D-AMP | 26.86 / 26.99 | 21.64 / 21.64 | 15.78 / 15.78 | 5.09 / 5.09 |
| | ReconNet | 25.59 / 26.22 | 22.63 / 23.23 | 20.27 / 21.06 | 17.63 / 18.30 |
| | DR²-Net | 28.73 / 29.10 | 23.94 / 24.30 | 21.16 / 21.85 | 18.01 / 18.41 |
| | CSRNet | 28.79 / 28.98 | 24.79 / 25.01 | 22.03 / 22.20 | 19.46 / 19.61 |
| | ASRNet | 29.61 / 29.80 | 27.68 / 27.85 | 24.52 / 24.67 | 21.87 / 22.01 |
| Boats | TVAL3 | 28.81 / 28.81 | 23.86 / 23.94 | 19.20 / 19.20 | 11.86 / 11.88 |
| | D-AMP | 29.26 / 29.26 | 21.90 / 21.87 | 16.01 / 16.01 | 5.34 / 5.34 |
| | ReconNet | 27.30 / 27.35 | 24.15 / 24.10 | 21.36 / 21.62 | 18.49 / 18.83 |
| | DR²-Net | 30.09 / 30.30 | 25.58 / 25.90 | 22.11 / 22.50 | 18.67 / 18.95 |
| | CSRNet | 30.14 / 30.36 | 25.65 / 25.80 | 22.27 / 22.44 | 18.94 / 19.01 |
| | ASRNet | 31.28 / 31.64 | 28.86 / 29.17 | 25.52 / 25.72 | 21.53 / 21.69 |
| Cameraman | TVAL3 | 25.69 / 25.70 | 21.91 / 21.92 | 18.30 / 18.33 | 11.97 / 12.00 |
| | D-AMP | 24.41 / 24.54 | 20.35 / 20.35 | 15.11 / 15.11 | 5.64 / 5.64 |
| | ReconNet | 23.15 / 23.59 | 21.28 / 21.66 | 19.26 / 19.72 | 17.11 / 17.49 |
| | DR²-Net | 25.62 / 25.90 | 22.46 / 22.74 | 19.84 / 20.30 | 17.08 / 17.34 |
| | CSRNet | 25.81 / 25.74 | 22.29 / 22.48 | 20.17 / 20.34 | 17.70 / 17.90 |
| | ASRNet | 26.46 / 22.66 | 25.00 / 25.13 | 22.74 / 22.88 | 19.77 / 19.89 |
| Foreman | TVAL3 | 35.42 / 35.54 | 28.69 / 28.74 | 20.63 / 20.65 | 10.97 / 11.01 |
| | D-AMP | 35.45 / 34.04 | 25.51 / 25.58 | 16.27 / 16.78 | 3.84 / 3.83 |
| | ReconNet | 29.47 / 30.78 | 27.09 / 28.59 | 23.72 / 24.60 | 20.04 / 20.33 |
| | DR²-Net | 33.53 / 34.28 | 29.20 / 30.18 | 25.34 / 26.33 | 20.59 / 21.08 |
| | CSRNet | 34.73 / 34.90 | 30.96 / 31.35 | 27.66 / 28.06 | 23.02 / 23.03 |
| | ASRNet | 35.85 / 36.19 | 33.79 / 34.09 | 30.56 / 30.78 | 25.77 / 26.14 |
| House | TVAL3 | 32.08 / 32.13 | 26.29 / 26.32 | 20.94 / 20.96 | 11.86 / 11.90 |
| | D-AMP | 33.64 / 32.68 | 24.55 / 24.53 | 16.91 / 17.37 | 5.00 / 5.02 |
| | ReconNet | 28.46 / 29.19 | 26.69 / 26.66 | 22.58 / 23.18 | 19.31 / 19.52 |
| | DR²-Net | 31.83 / 32.52 | 27.53 / 28.40 | 23.92 / 24.70 | 19.61 / 19.99 |
| | CSRNet | 32.33 / 32.75 | 28.24 / 28.68 | 24.40 / 24.69 | 20.45 / 20.78 |
| | ASRNet | 33.44 / 33.84 | 31.47 / 31.87 | 27.82 / 28.21 | 23.13 / 23.31 |
| Peppers | TVAL3 | 29.62 / 29.65 | 22.64 / 22.65 | 18.21 / 18.22 | 11.35 / 11.36 |
| | D-AMP | 29.84 / 28.58 | 21.39 / 21.37 | 16.13 / 16.46 | 5.79 / 5.85 |
| | ReconNet | 24.77 / 25.16 | 22.15 / 22.67 | 19.56 / 20.00 | 16.82 / 16.96 |
| | DR²-Net | 28.49 / 29.10 | 23.73 / 24.28 | 20.32 / 20.78 | 16.90 / 17.11 |
| | CSRNet | 28.42 / 28.54 | 24.35 / 24.65 | 21.18 / 21.41 | 17.57 / 17.62 |
| | ASRNet | 29.72 / 30.18 | 27.03 / 27.37 | 24.03 / 24.32 | 20.17 / 20.33 |
| Mean PSNR | TVAL3 | 27.84 / 27.87 | 22.84 / 22.86 | 18.39 / 18.40 | 11.31 / 11.34 |
| | D-AMP | 28.17 / 27.67 | 21.14 / 21.09 | 15.49 / 15.67 | 5.19 / 5.23 |
| | ReconNet | 25.54 / 25.92 | 22.68 / 23.23 | 19.99 / 20.44 | 17.27 / 17.55 |
| | DR²-Net | 28.66 / 29.06 | 24.32 / 24.71 | 20.80 / 21.29 | 17.44 / 17.80 |
| | CSRNet | 28.76 / 28.93 | 24.55 / 24.81 | 21.45 / 21.66 | 18.19 / 18.30 |
| | ASRNet | 29.85 / 30.17 | 27.65 / 27.96 | 24.40 / 24.65 | 20.51 / 20.66 |

### 3.2.2 Time complexity

The time complexity is a key factor for image compressive sensing. In the progress of reconstruction, the network-based algorithms are much faster than traditional iterative reconstruction methods, so we only compare the time complexity with ReconNet and DR²-Net. **Table 2** shows the average time for reconstructing nine sized 256×256 images of those network-based methods.

From the **Table 1** and **Table 2**, we can observe that the proposed CSRNet and ASRNet outperform the ReconNet and DR²-Net in terms of PSNR and time complexity. And our ASRNet obtains the best performance in the PSNR values and time complexity. Notably, ASRNet run fastest which is very important for real time applications.

Table 2. Time(in seconds) for reconstruction a single 256×256 image

| Models | MR=0.25 | MR=0.10 | MR=0.04 | MR=0.01 |
|---|---|---|---|---|
| ReconNet | 0.5376 | 0.5366 | 0.5415 | 0.5508 |
| DR² -Net | 1.2879 | 1.2964 | 1.3263 | 1.3263 |
| CSRNet | 0.5355 | 0.5375 | 0.5464 | 0.5366 |
| ASRNet | 0.3038 | 0.2984 | 0.3005 | 0.3025 |

### 3.2.3 Visual quality comparisons

Our proposed algorithm is compared with four representative CS recovery methods, TVAL3, D-AMP, ReconNet and DR²-Net in visual. **Fig.5-6** show the visual comparisons of *Parrots* in the case of measurement rate=0.1 with and without BM3D respectively. It is obvious that the proposed CSRNet and ASRNet are able to reconstruct more details and sharper, which offers better visual reconstruction results than other network-based algorithms. The other three groups are shown in **Fig.7-12**. The test images under different rates with or without BM3D are shown in the **Fig.13-16**. We can see that our proposed CSRNet and ASRNet outperforms ReconNet and DR²-Net at every measurement rates.

## 5. Conclusion

In this paper, two cascaded reconstructed networks are proposed for different CS sampling methods. In most previous works, the sample matrix is a random matrix in CS process. And the first network is a compatibly sampling reconstruction network (CSRNet), which can reconstruct high-quality images from its compressively sensed measurement sampled by a traditional random matrix. The second network is adaptively sampling reconstruction network (ASRNet), by matching automatically sampling module with corresponding residual reconstruction module. And the sampling module could perfectly solve the problem of sampling efficiency in compressive sensing. Experimental results show that the proposed networks, CSRNet and ASRNet, have achieved the significant improvements in reconstruction results over the traditional and neural network based CS reconstruction algorithms both in terms in quality and time complexity. Furthermore, ASRNet can achieve more than 1 dB gain, as compared with CSRNet.

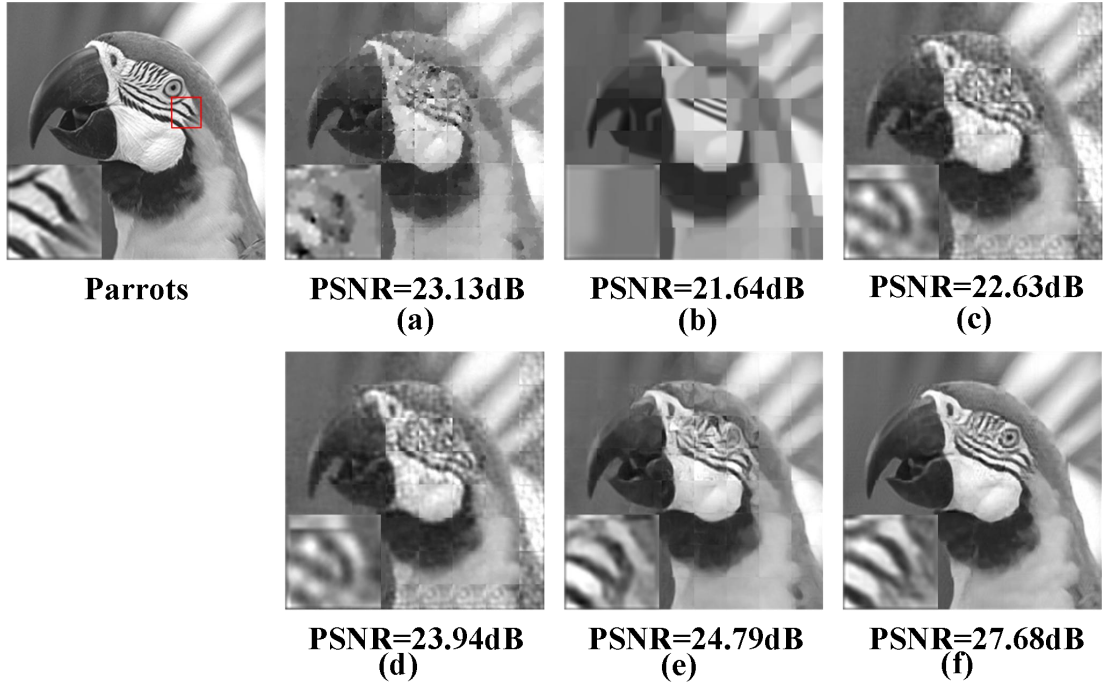

**Fig.5.** The reconstruction results of *Parrots* at the measurement rate of 0.1 without BM3D, (a)TVAL3; (b) DAMP; (c) ReconNet; (d) DR$^2$-Net; (e) CSRNet; (f) ASRNet.

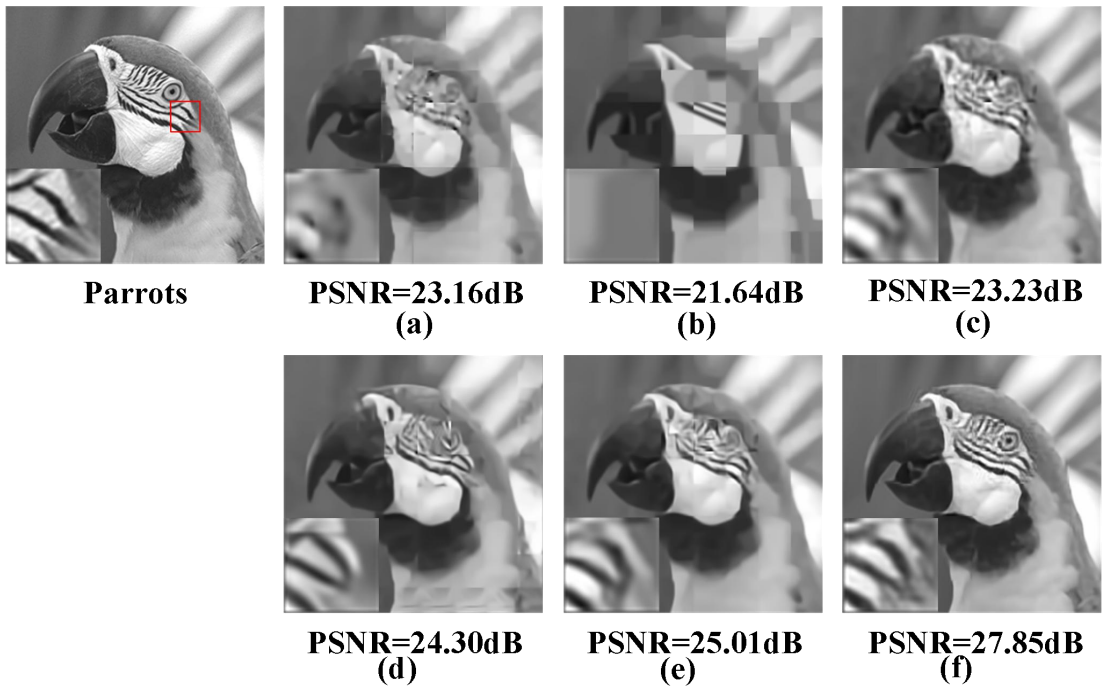

**Fig.6.** The reconstruction results of *Parrots* at the measurement rate of 0.1 with BM3D, (a)TVAL3; (b) DAMP; (c) ReconNet; (d) DR$^2$-Net; (e) CSRNet; (f) ASRNet..

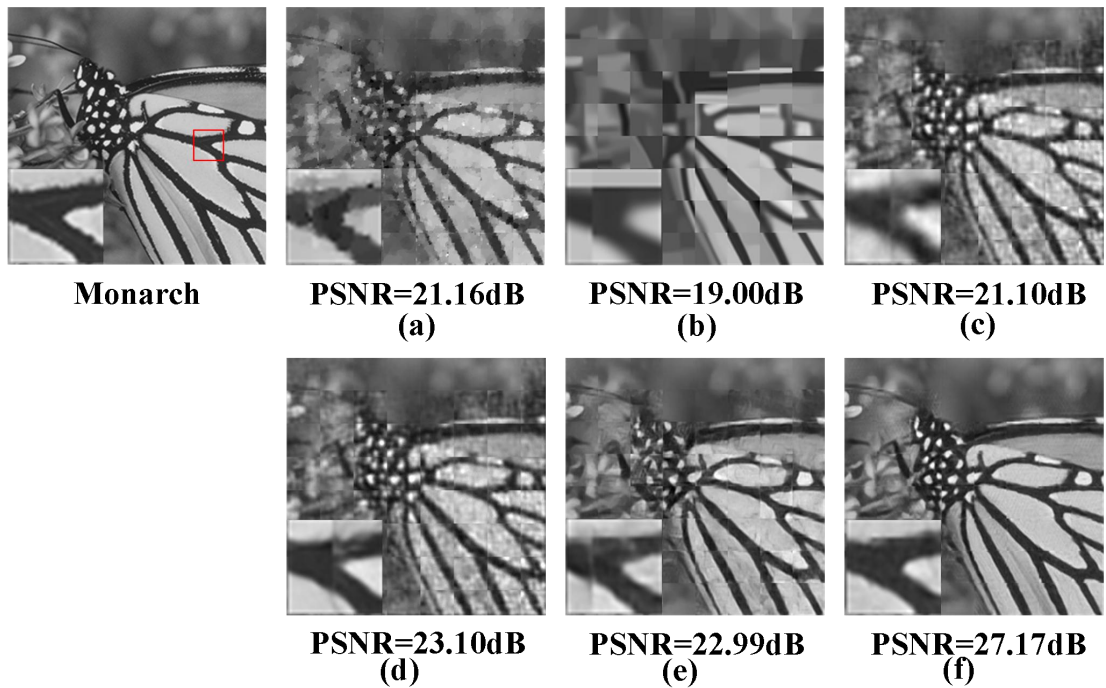

**Fig.7.** The reconstruction results of *Monarch* at the measurement rate of 0.1 without BM3D, (a)TVAL3; (b) DAMP; (c) ReconNet; (d) DR² -Net; (e) CSRNet; (f) ASRNet..

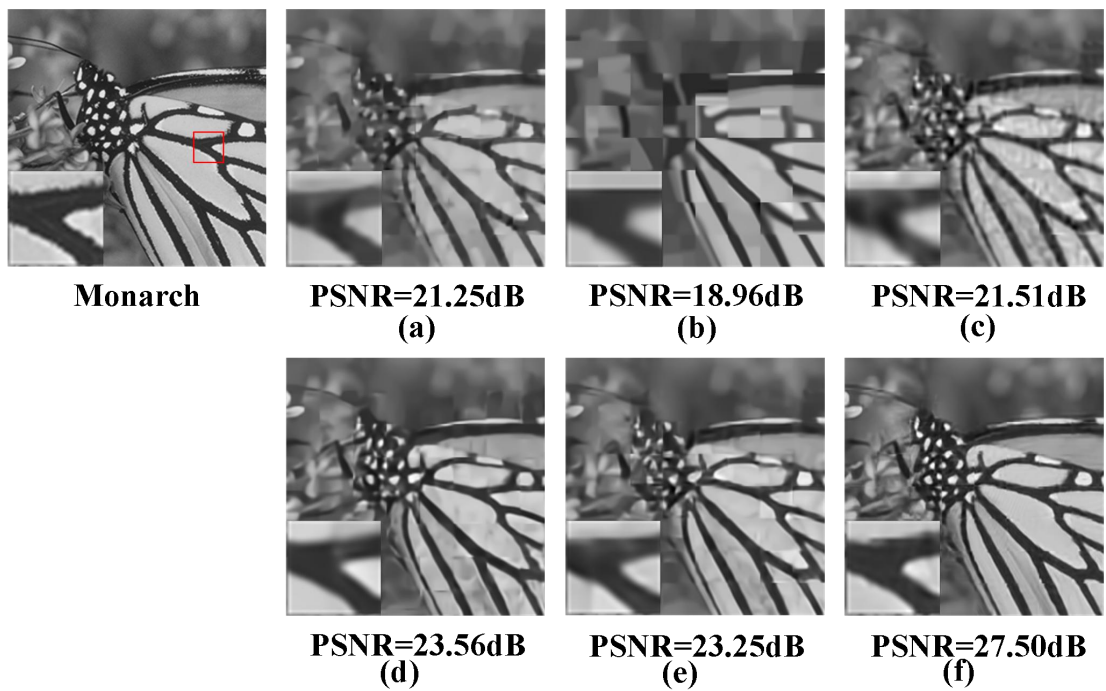

**Fig.8.** The reconstruction results *Monarch* at the measurement rate of 0.1 with BM3D, (a)TVAL3; (b) DAMP; (c) ReconNet; (d) DR² -Net; (e) CSRNet; (f) ASRNet.

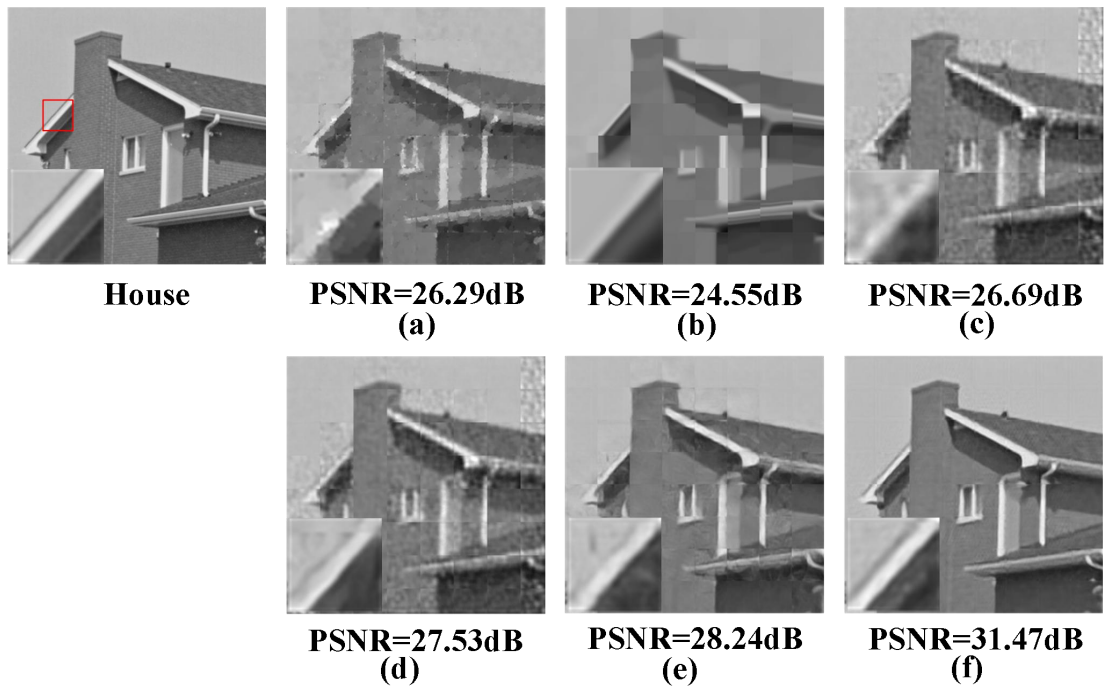

**Fig.9.** The reconstruction results of *House* at the measurement rate of 0.1 without BM3D, (a)TVAL3; (b) DAMP; (c) ReconNet; (d) DR² -Net; (e) CSRNet; (f) ASRNet..

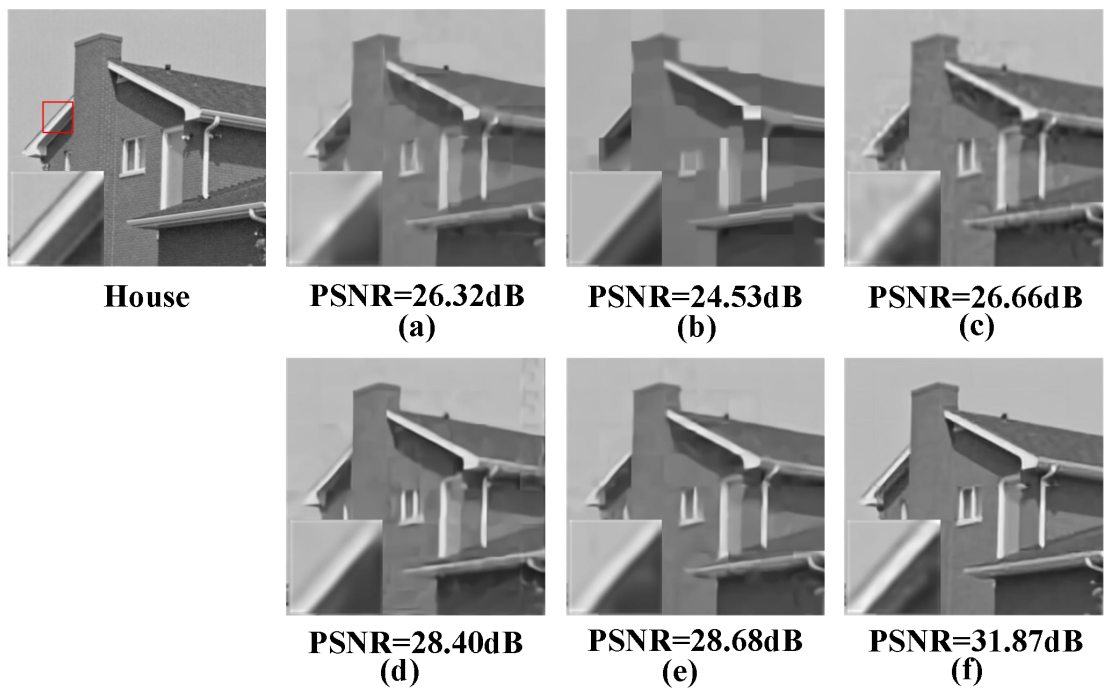

**Fig.10.** The reconstruction results of *House* at the measurement rate of 0.1 with BM3D, (a)TVAL3; (b) DAMP; (c) ReconNet; (d) DR² -Net; (e) CSRNet; (f) ASRNet..

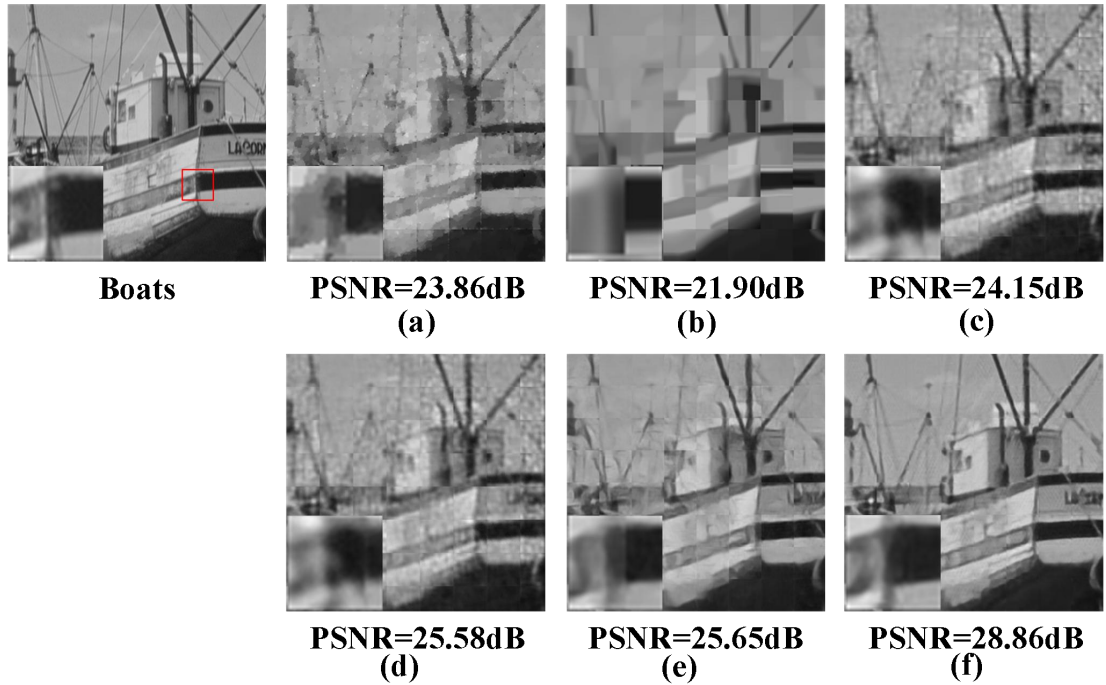

**Fig.11.** The reconstruction results of *Boats* at the measurement rate of 0.1 without BM3D, (a)TVAL3; (b) DAMP; (c) ReconNet; (d) DR² -Net; (e) CSRNet; (f) ASRNet..

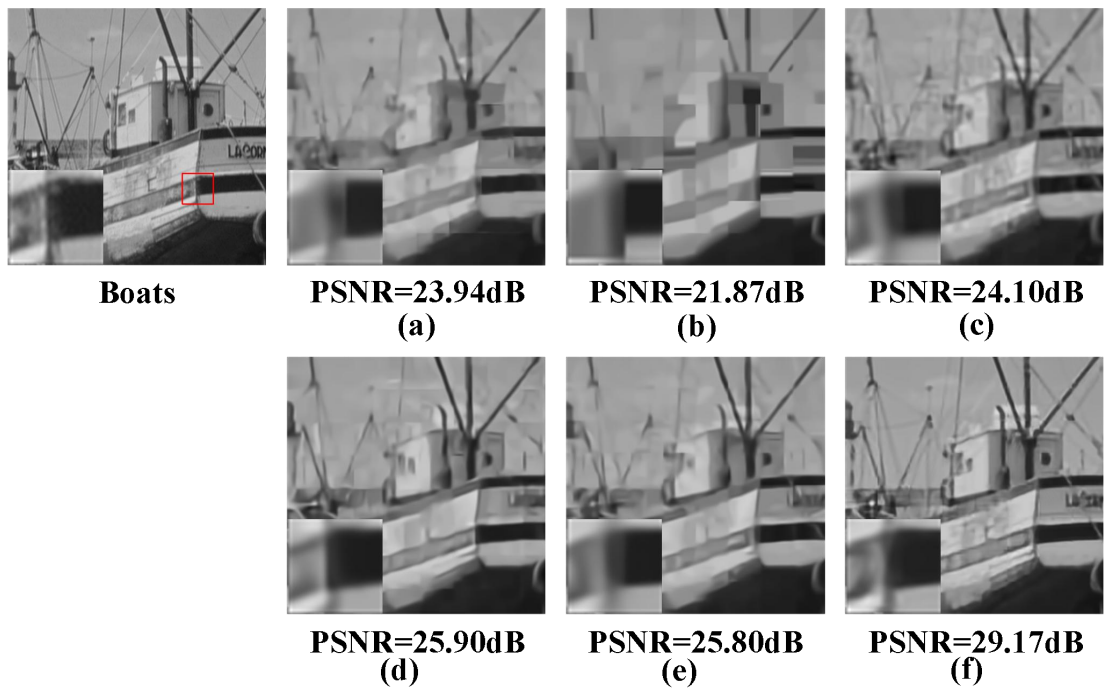

**Fig.12.** The reconstruction results of *Boats* at the measurement rate of 0.1 with BM3D, (a)TVAL3; (b) DAMP; (c) ReconNet; (d) DR² -Net; (e) CSRNet; (f) ASRNet..

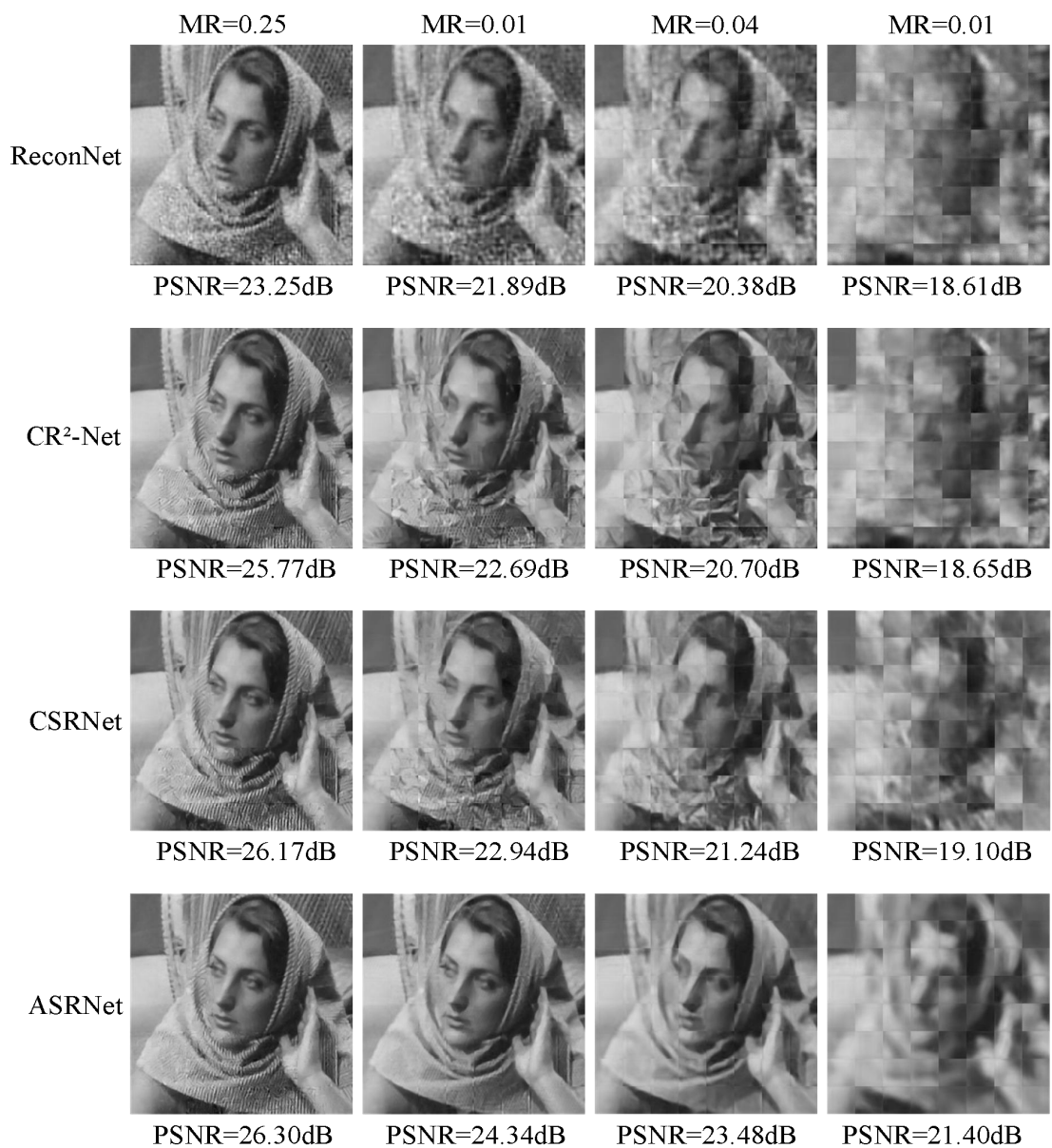

Fig.13.The reconstruction results *Barbara* at the measurement rate of 0.25, 0.1, 0.04 and 0.01 without BM3D.

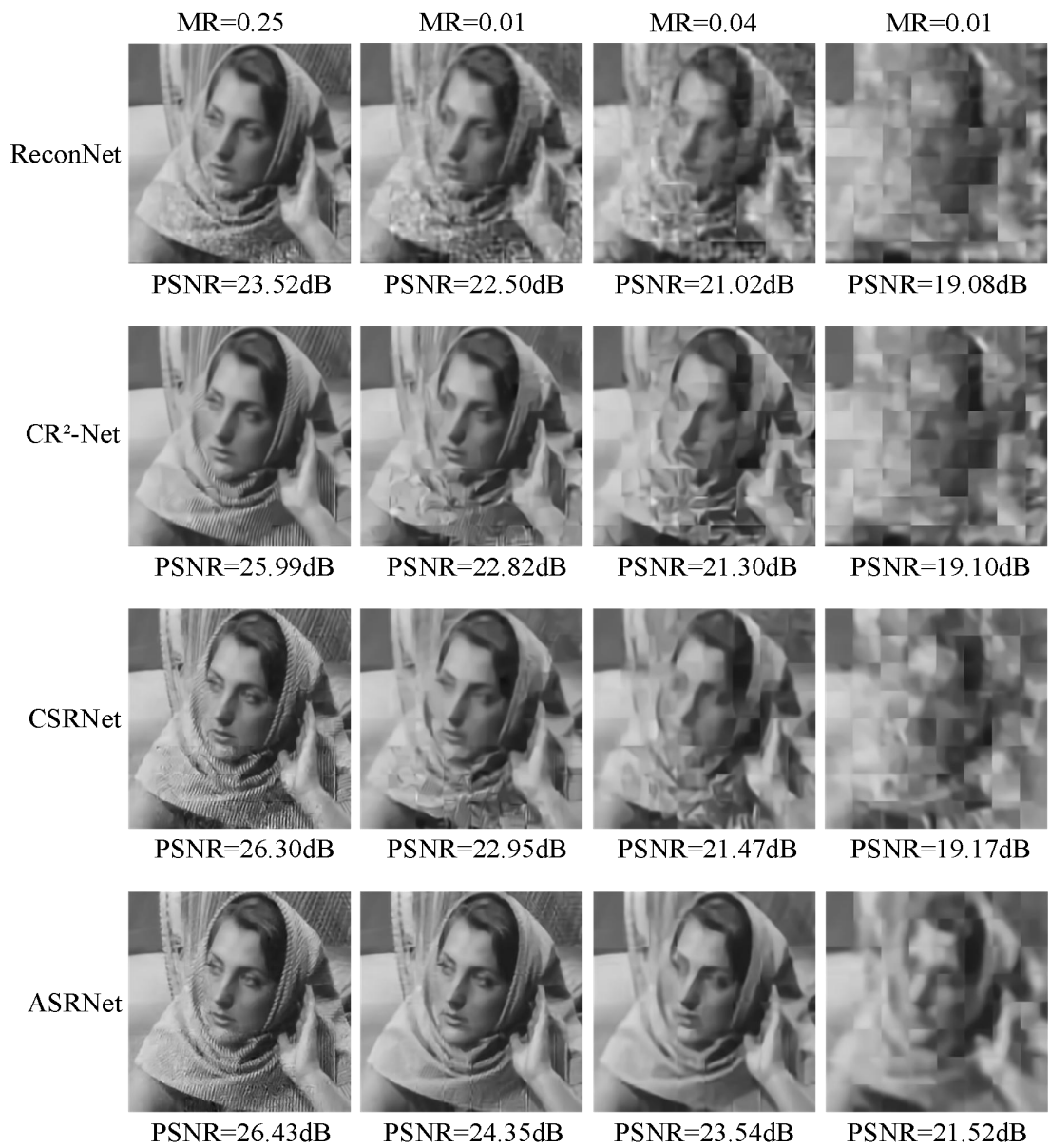

Fig.14. The reconstruction results *Barbara* at the measurement rate of 0.25, 0.1, 0.04 and 0.01 with BM3D.

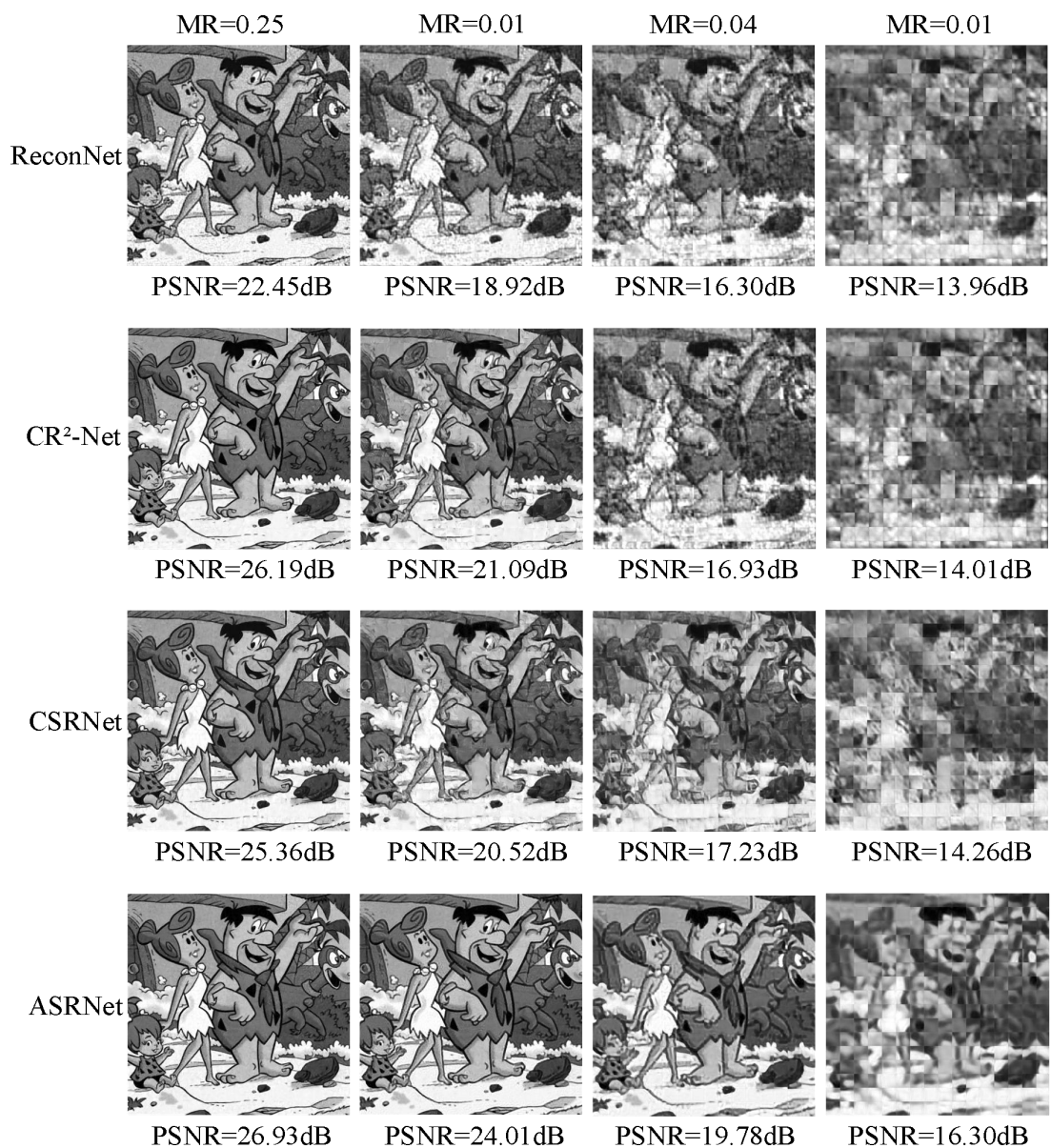

Fig.15. The reconstruction results *Flinstones* at the measurement rate of 0.25, 0.1, 0.04 and 0.01 without BM3D.

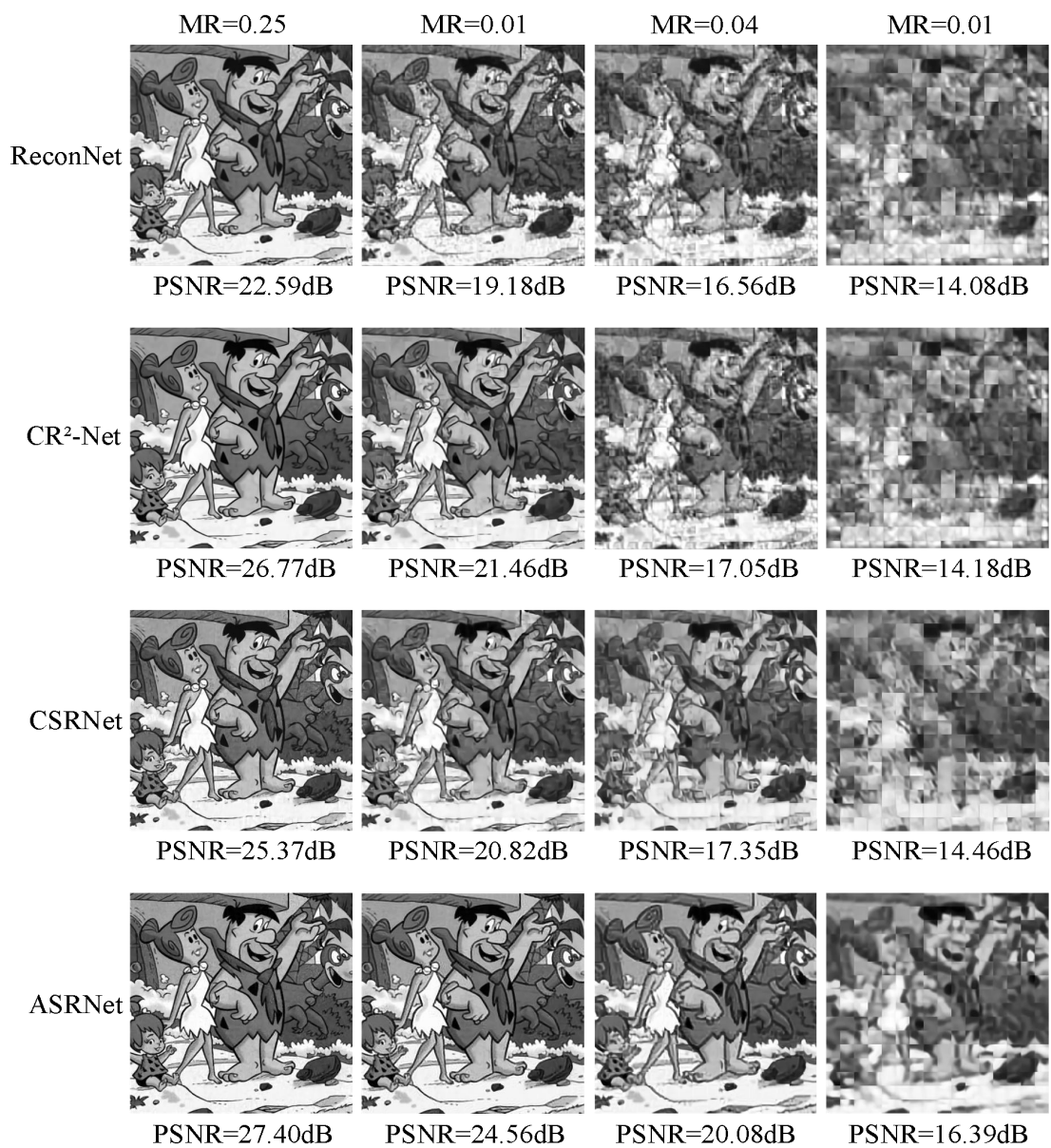

**Fig.16.** The reconstruction results *Flinstones* at the measurement rate of 0.25, 0.1, 0.04 and 0.01 with BM3D.